\documentclass[twocolumn]{aastex63}
\pdfoutput=1

\shorttitle{Galaxy Alignments with Surrounding Structure}
\shortauthors{Desai \& Ryden}
\graphicspath{{./}{figures/}}

\begin{document}

\title{Galaxy Alignments with Surrounding Structure in the Sloan Digital Sky Survey}


\correspondingauthor{Barbara S. Ryden}
\email{ryden.1@osu.edu}

\author[0000-0002-2164-859X]{Dhvanil D. Desai}
\affiliation{Department of Astronomy,
The Ohio State University,
140 W. $18^{th}$ Ave.,
Columbus, OH 43210 USA}
\affiliation{Institute for Astronomy,
University of Hawaii,
2680 Woodlawn Drive,
Honolulu, HI 96822 USA}

\author[0000-0001-8164-8703]{Barbara S. Ryden}
\affiliation{Department of Astronomy,
The Ohio State University,
140 W. $18^{th}$ Ave.,
Columbus, OH 43210 USA}
\affiliation{Center for Cosmology \& Astroparticle Physics,
The Ohio State University,
191 W. Woodruff Ave.,
Columbus, OH 43210 USA}


\begin{abstract}

\noindent
Using data from the Sloan Digital Sky Survey (SDSS) Legacy Survey, we study
the alignment of luminous galaxies with spectroscopic data with the
surrounding larger-scale structure as defined by galaxies with only photometric
data. We find that galaxies from the red sequence have a statistically
significant tendency for their apparent long axes to align parallel to the projected
surrounding structure. Red galaxies more luminous than the median of our sample
($M_r < -21.78$) have a mean alignment angle $\langle \Phi \rangle < 45\degr$,
indicating preferred parallel alignment, at a significance level $>4.5 \sigma$
on projected scales $0.1\,\mathrm{Mpc} < r_p \leq 7.5$\,Mpc. Fainter red
galaxies have $\langle \Phi \rangle < 45\degr$ at a significance level $>4.3\sigma$
at scales $1\,\mathrm{Mpc} < r_p < 3$\,Mpc. At a projected
scale $r_p = 3.0$\,Mpc, the mean alignment angle decreases steadily with
increasing luminosity for red galaxies with $M_r \lesssim -22.5$, reaching
$\langle \Phi \rangle = 40.49\degr \pm 0.56\degr$ for the most luminous one percent
($M_r \sim -23.57$). Galaxies from the blue sequence show no statistically significant
tendency for their axes to align with larger-scale structure, regardless of galaxy
luminosity. Galaxies in higher-density regions do not show a statistically significant
difference in mean alignment angle from galaxies in lower-density regions; this holds
true for the faint blue, luminous blue, faint red, and luminous red subsets.

\end{abstract}



\section{Introduction and background}
\label{sec:intro}

The study of the intrinsic alignment of galaxies with larger scale structure has
a long history 
\citep{1939MNRAS..99..534B,1955AJ.....60..415W,1958PASP...70..461R,1964MNRAS.127..517B}.
Statistical analysis of these intrinsic alignments can shed light on how the
large-scale structure of the universe affects the formation and evolution of galaxies.
Luminous late-type galaxies, for example, have rotationally supported disks.
Tidal torque theory \citep{1951pca..conf..195H,1969ApJ...155..393P} explains
how disk galaxies are spun up through the interaction of a protogalaxy's
quadrupole moment with the tidal field of the surrounding matter distribution.
Thus, it is reasonable to assume some sort of alignment between the disk's angular
momentum vector and the surrounding large-scale structure. In addition, the
outer regions of early-type galaxies will also be shaped by the tidal gravitational
field of surrounding matter \citep{2001MNRAS.320L...7C,2019JCAP...04..031R}.
The predictions of tidal torque theory can be extended into the nonlinear regime
by analytic methods \citep{2001MNRAS.320L...7C,2007NJPh....9..444B,2015JCAP...08..015B,2022ApJ...926..200E}.

The use of numerical simulations to study the evolution of intrinsic alignments
allows the introduction of non-gravitational effects. Generally, simulations
show that more massive and more luminous galaxies show stronger alignment
signals \citep{2017MNRAS.468..790H,2019MNRAS.487.1607G,2021MNRAS.501.5859T}.
The evolution of intrinsic alignment in numerical simulations is found to
depend on many factors \citep{2020MNRAS.491.4116B}. In general, high-mass
galaxies have an alignment signal that increases from $z \sim 1$ to $z \sim 0$
\citep{2020arXiv201007951Z,2020arXiv200910735S}; however, low-mass galaxies
tend to have an alignment that decreases with time \citep{2018MNRAS.481.4753C}.
Numerical simulations also permit a full three-dimensional view of how galaxies
are oriented with respect to different structures in the cosmic web. For
instance, \citet{2018MNRAS.481.4753C} find that the spin axis of low-mass
galaxies aligns with filaments in the cosmic web, while the spin axis of higher-mass
galaxies aligns perpendicular to filaments.

Observational studies of intrinsic alignment generally require using data
from large surveys, given the weakness of the alignment signal. Within the
Local Group, for instance, alignments are difficult to detect with high
statistical significance, even with ultra-faint dwarfs added to the sample
\citep{2017MNRAS.472.2670S}. However, a deeper study of $N \sim 6000$ systems
similar to the Local Group, dominated by two luminous galaxies, reveals that
satellites between the two spirals tend to have their projected long axis
perpendicular to the line connecting the bright primary galaxies, while outer
satellites tend to be radially aligned \citep{2019MNRAS.484.4325W}. In larger
groups containing a single bright galaxy, satellite galaxies tend to align with
the apparent long axis of the brightest group galaxy, with red satellites showing
a stronger alignment signal than blue satellites \citep{2019A&A...628A..31G}.
In addition to alignments between satellite galaxies and the group or cluster in
which they are found, alignments are also seen  between the satellite distribution
and filaments in the cosmic web \citep{2020ApJ...900..129W}. Several studies,
reviewed by \citet{2015SSRv..193....1J}, indicate that disk galaxies tend to
align their spin perpendicular to the direction of filaments in the large-scale
structure. However, a more recent MaNGA survey \citep{2021MNRAS.504.4626K} indicates
that while S0 galaxies tend to have a spin axis perpendicular to the nearest
filament, spirals of later type tend to have their spin axis parallel to the nearest
filament. In addition, the alignment signal for late-type galaxies is weaker than
the signal for early-type galaxies, which tend to align their apparent major axes
parallel to the direction of filaments \citep{2015SSRv..193....1J,2019MNRAS.485.2492C}.

Observational studies emphasize that red, early-type galaxies tend to have a stronger
alignment than blue, late-type galaxies \citep{2015SSRv..193....1J,2019A&A...624A..30J}.
\citet{2022ApJ...924L...3T} find that red galaxies show detectable alignment at
redshifts as high as $z \sim 1.3$, at the $2.5\sigma$ level. Among the population
of red galaxies, the most luminous galaxies show the strongest alignment signal
\citep{2021A&A...654A..76F,2022arXiv220400683R}.

The study of intrinsic alignments, in addition to giving insight into the
formation and evolution of large-scale structure, is essential for interpreting
the results of weak gravitational lensing measurements
\citep{1967ApJ...150..737G,2009ApJ...694L..83O}.
Weak lensing by the intervening mass distribution produces a shear distortion in
a distant galaxy's image. Although intrinsic alignment is sometimes described as
a ``contaminant'' of the image alignment caused by weak gravitational lensing,
the same tidal shear fields that create weak lensing distortions can also physically
distort the outer regions of a galaxy \citep{2020ApJ...899L...5P,2021MNRAS.505.2594G}.
By studying galaxies at low redshift ($z \lesssim 0.1$), where the weak lensing
effect is negligible compared to intrinsic alignments, we can quantify the
intrinsic alignment signal on its own. Given the known dependence of intrinsic
alignment on galaxy color and luminosity, we will be careful to distinguish
between red, early-type galaxies and blue, late-type galaxies in our study of
galaxy alignment with surrounding structure, and will pay particular attention
to the dependence of the alignment on the galaxy's luminosity.

\section{Data and definitions}
\label{sec:data}

Our study used galaxy data from the Legacy Survey of the Sloan Digital Sky
Survey (SDSS) \citep{2000AJ....120.1579Y,2006AJ....131.2332G}. Data were
downloaded from Data Release 16 of the SDSS
\citep{2020ApJS..249....3A}. The SDSS Legacy imaging survey covered
14{,}555 deg$^2$, recording imaging data for $\sim$50 million galaxies;
the SDSS Legacy spectroscopic survey provided spectroscopic data for
$\sim$1.5 million of those galaxies \citep{2011AJ....142...72E}.
The SDSS database has proved highly valuable for studies of the
alignment of galaxy images with larger scale structure
\citep{2007ApJ...670L...1L,2007MNRAS.381.1197H,2008MNRAS.385.1511W,2008MNRAS.389.1127P,2009ApJ...694..214O,2010MNRAS.408..897J,2013ApJ...779..160Z,2015ApJ...798...17Z,2017AA...599A..31H,2018ApJ...859..115W,2020ApJ...900..129W,2020MNRAS.500.1895Z}.
The footprint of the SDSS Legacy survey is sufficiently large to allow
the detection of alignments on scales of $\sim$10\,Mpc or larger
\citep{2012MNRAS.423..856S,2019MNRAS.485.2492C}. In this paper, we use
SDSS Legacy data to study the alignment of ``target galaxies,''
drawn from the spectroscopic survey and thus having accurate spectroscopic
redshifts, with the larger scale structure defined by ``surrounding
galaxies,'' drawn from the imaging survey and thus having only imprecise
photometric redshifts.

The SDSS imaging survey used five broadband filters $u$, $g$, $r$, $i$, 
and $z$ \citep{1996AJ....111.1748F,2010AJ....139.1628D}. Photometric
parameters in the $r$ band (effective wavelength $\lambda_r = 6261$\AA )
were used to determine the location, axis ratio, and orientation of the
galaxies in our sample. Galaxy axis ratios and orientation can be
wavelength-dependent \citep{2019A&A...622A..90G}. Since galaxies tend to
be bluer with increasing radius, choosing the $r$ band, rather than a
shorter-wavelength band, de-emphasizes the contributions of the outer
regions of target galaxies. The $u-r$ color index was used to
define the color of galaxies. The $u$ band, with effective wavelength
$\lambda_u = 3557$\AA , is sensitive to the presence of hot
stars; thus the $u-r$ color index is a good diagnostic of the
presence of recent star formation. From the SDSS table \texttt{PhotoObj}, 
following the suggestion of \citet{2005astro.ph..8564S}, 
\texttt{cModelMag\_r} was used for the absolute $r$ band magnitude and 
\texttt{modelMag\_r} and \texttt{modelMag\_u} were used for the $u-r$ color 
index.\footnote{We find that using Petrosian magnitudes (\texttt{petroMag})
instead of Model (\texttt{modelMag}) and Composite Model magnitudes 
(\texttt{cModelMag}) for color and absolute magnitude does 
not change the qualitative conclusions of this paper.}

The SDSS spectroscopic survey, from which we drew our target galaxies,
is complete to a limiting $r$ band magnitude $r = 17.77$
\citep{2002AJ....124.1810S}.
We selected target galaxies with spectroscopic redshifts in the
range $0.02 < z < 0.25$. The lower redshift limit eliminates galaxies
whose peculiar motion contributes significantly to the redshift. The
higher redshift limit mostly excludes galaxies that form the
Luminous Red Galaxy (LRG) sample \citep{2001AJ....122.2267E}.
The excluded higher redshift galaxies have a
median absolute magnitude $M_r = -22.69$ and median color $u-r = 3.08$.
Adding these LRGs to our sample would not greatly contribute to our goal of 
identifying how alignment properties differ between the red sequence and 
the blue sequence of galaxies, and by eliminating them from our sample, 
we reduce possible effects of evolution ($z = 0.25$ corresponds to a 
lookback time $t \approx 3.0\,\mathrm{Gyr}$) and of weak lensing shear. 

We compute distance modulus as a function of spectroscopic redshift $\mu (z)$ 
using the Python package \texttt{astropy.cosmology} 
\citep{astropy:2013,astropy:2018}, assuming a flat $\Lambda$CDM cosmology 
with parameters from the \textit{Planck} 2018 results 
\citep{2020A&A...641A...6P}. In particular, this assumes 
$H_0 = 67.4\,\mathrm{km\,s}^{-1}\,\mathrm{Mpc}^{-1}$ 
and $\Omega_{m,0} = 0.315$.
We corrected the $u$ and $r$ band magnitudes for Galactic extinction
exactly as done by the SDSS pipeline; this correction uses the dust map
of \citet{1998ApJ...500..525S}, and assumes a 7000\,K source and a ratio
of total to selective extinction $R_V = 3.1$ \citep{2011ApJ...737..103S}.
For each extragalactic source, the correction yields an estimated
Galactic extinction $A_{u,Gal}$ and $A_{r,Gal}$. The $K$-correction
for each source \citep{2002astro.ph.10394H} was done using the code
of \citet{2007AJ....133..734B}. The absolute $r$ band magnitude is then
\begin{equation}
    M_r = r - \mu (z) - A_{r,Gal} - K_r (z) 
    \label{abs_mag}
\end{equation}
and the corrected $u-r$ color is
\begin{equation}
    (u-r)_\mathrm{corr} = u - r - A_{u,Gal} + A_{r,Gal} - K_u (z) + K_r (z)  .
\end{equation}

\begin{figure}[ht!]
\epsscale{1.1788}
\plotone{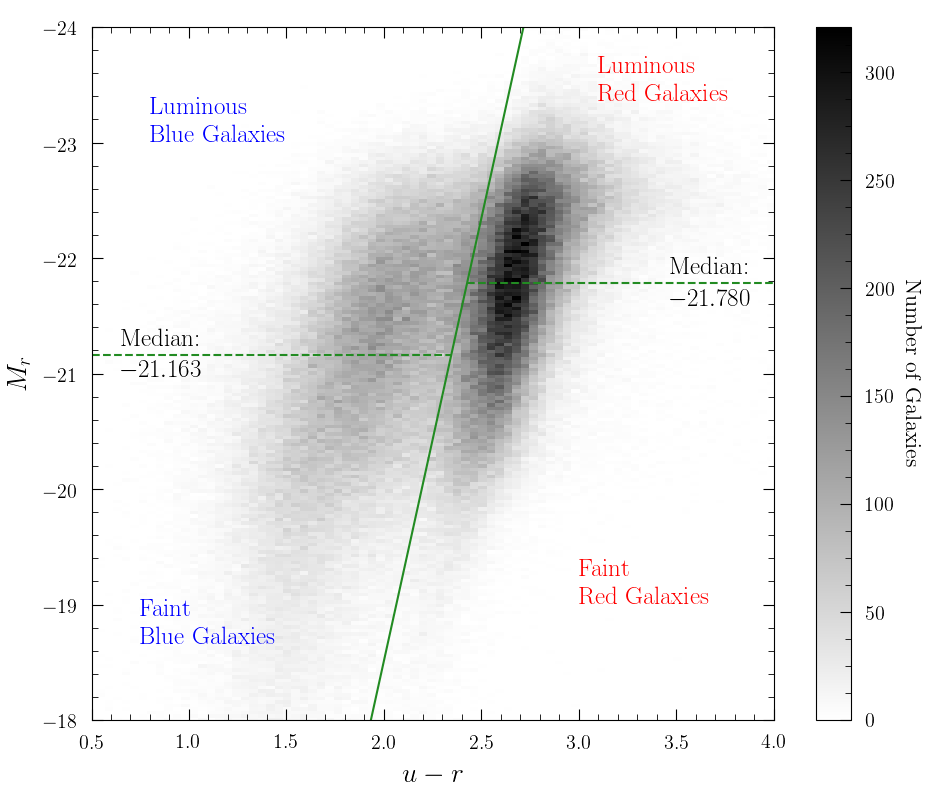}
\caption{Color--magnitude diagram ($u-r$ versus $M_r$) for the
target galaxies. The four subsamples (luminous blue, luminous red,
faint blue, and faint red) are labeled. Solid green line is the color
divider of \citet{James_2022}.}
\label{fig:color_mag}
\end{figure}

Our final sample consists of 396{,}718 target galaxies, with median 
redshift $z_\mathrm{med} = 0.107$. We divide our full sample of target 
galaxies into a blue sequence and a red sequence using the color divider 
of \citet{James_2022}:\footnote{We find that using the color divider
of \citet{2004ApJ...600..681B}, based on a shallower sample of SDSS
Legacy galaxies, does not change the qualitative conclusions of this paper.}
\begin{equation}
u-r = 2.391 - 0.1305 (M_r + 21.5) .
\label{eq:color_divider}
\end{equation}
With this definition, 184{,}788 target galaxies lie on the blue side
of the divider and 211{,}930 lie on the red side. Finally, we divided
the blue galaxies and the red galaxies into a luminous subsample and
a faint subsample at their respective medians: $M_r = -21.163$ for the
blue target galaxies, and $M_r = -21.780$ for the red target galaxies.
The interquartile range for the blue galaxies is 
$-21.86 < M_r < -20.32$ and for the red galaxies is 
$-22.35 < M_r < -21.09$. Figure~\ref{fig:color_mag} shows the 
extinction and $K$-corrected color--magnitude diagram for our sample 
of 396{,}718 target galaxies; the \citet{James_2022} color divider is 
plotted as the solid green line.

\begin{figure*}[ht!]
\epsscale{0.85}
\plotone{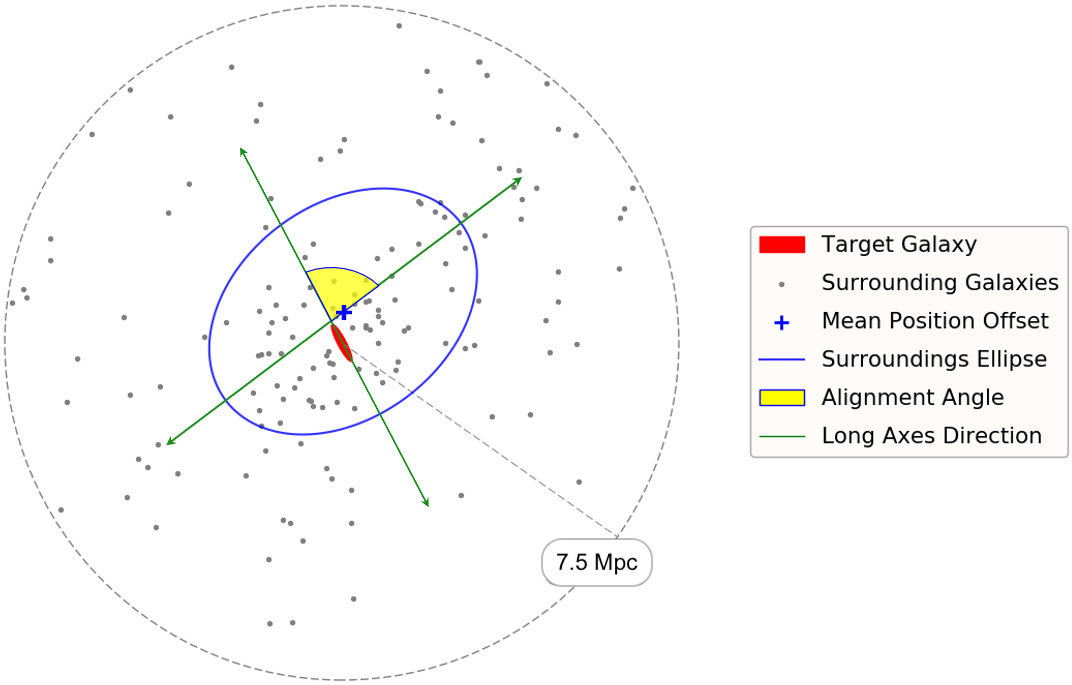}
\caption{Definition of the alignment angle $\Phi$ of the target
galaxy (filled red ellipse) relative to the distribution of
surrounding galaxies (fitted by the open blue ellipse).}
\label{fig:align_define}
\end{figure*}

For each target galaxy, the position angle $\phi_\mathrm{pa}$ is taken from
the adaptive moments parameters in the SDSS Legacy database; the phase of the
position angle is chosen so that $0\degr \leq \phi_\mathrm{pa} < 180\degr$,
running from north through east.
Each target galaxy drawn from the SDSS Legacy spectroscopic survey has
associated ``surrounding galaxies.'' These surrounding galaxies are
defined as galaxies from the SDSS Legacy photometric survey that
lie within a projected distance $r_p$ (computed using \texttt{astropy}) of 
the target galaxy, at the target galaxy's spectroscopic redshift $z$ given the 
angular separation $\theta$ between the target galaxy and the surrounding 
galaxy.
We choose $r_p = 3.0\, \mathrm{Mpc}$ as our primary separation for analyses, 
but we also consider the broader range of $r_p = 0.1 \to 7.5\, \mathrm{Mpc}$.
We selected our target galaxies
from an area embedded within the main photometric survey footprint, with
a buffer of width $\sim 5\degr$ between the selection area and the
survey boundary. This ensures that even for the lowest-redshift target
galaxies, none of the surrounding galaxies fall outside the photometric 
survey footprint.

Since the majority of the surrounding galaxies have photometric
data only, they have only photometric redshifts, $z_\mathrm{phot}$,
rather than the more accurate spectroscopic redshifts of the target
galaxies \citep{2016MNRAS.460.1371B}. To eliminate surrounding galaxies
with a high probability of being foreground or background contaminants,
we impose the additional constraint that a surrounding galaxy
must have $|z - z_\mathrm{phot}| < 2 \delta z_\mathrm{phot}$,
where $z$ is the spectroscopic redshift of the target galaxy and
$\delta z_\mathrm{phot}$ is the rms error in $z_\mathrm{phot}$ for the
surrounding galaxy \citep{2016MNRAS.460.1371B}. We impose the additional
constraint that surrounding galaxies must have $\delta z_\mathrm{phot} < 0.04$.
Eliminating in this way the galaxies that are likely to
be foreground or background galaxies, the median number of surrounding
galaxies per target galaxy for a projected separation of 
$r_p = 3.0\,\mathrm{Mpc}$ ranges from $N_\mathrm{med} \approx 500$
for the lowest-redshift target galaxies to $N_\mathrm{med} \approx 40$
for the highest-redshift target galaxies. We also limit our sample to 
target galaxies with $N \geq 4$ surrounding galaxies, to avoid statistically
unreliable alignments; this cut eliminates only 8 target galaxies. For the entire
sample of target galaxies, the mean number of surrounding galaxies within a
projected distance $r_p = 3.0\, \mathrm{Mpc}$ is $\overline{N} = 177$.

Studies of the alignment of galaxies with surrounding structure
have used multiple definitions of the alignment angle $\Phi$
\citep{2015SSRv..193....1J}; it is thus important to clearly
describe our own definition. A target galaxy is at right ascension
$\alpha_t$ and declination $\delta_t$. It is surrounded by a
population of $N$ surrounding galaxies that survive our cuts in
projected separation and photometric redshift. The $i^\mathrm{th}$
surrounding galaxy has right ascension $\alpha_i$ and declination
$\delta_i$. Since the surrounding galaxies are at small angular
separation from the target galaxy, we may safely use the ``flat
celestial sphere'' approximation, and compute the coordinates
of the surrounding galaxies in a Cartesian system whose origin is
at the position of the target galaxy. In this system, the $x$ axis
is in the north -- south direction, with $x$ increasing northward,
while the $y$ axis is in the east -- west direction, with $y$
increasing eastward. The position of each surrounding galaxy in
this system is
\begin{equation}
x_i = \delta_i - \delta_t \label{eq:x_position} , \qquad
y_i = (\alpha_i - \alpha_t ) \cos ( [\delta_i+\delta_t] / 2 )  \label{eq:y_position} .
\end{equation}
Weighing each galaxy equally, the mean offset of the surrounding
galaxies from the target galaxy (blue cross in Figure~\ref{fig:align_define}) 
is given by the first order moments
\begin{equation}
\mu_x = \frac{1}{N} \sum_{i=1}^N x_i , \label{eq:mu_x} \qquad
\mu_y = \frac{1}{N} \sum_{i=1}^N y_i  \label{eq:mu_y} .
\end{equation}
The second order moments are then
\begin{eqnarray}
\mu_{xx} &=& \frac{1}{N} \sum_{i=1}^N (x_i - \mu_x)^2 , \quad \label{eq:mu_xx}
\mu_{yy} = \frac{1}{N} \sum_{i=1}^N (y_i - \mu_y)^2 , \label{eq:mu_yy} \nonumber \\
\mu_{xy} &=& \frac{1}{N} \sum_{i=1}^N
 (x_i - \mu_x)(y_i - \mu_y)  \label{eq:mu_xy} .
\end{eqnarray}
The shape of the distribution of surrounding galaxies can then be approximated as an
ellipse (open blue ellipse in Figure~\ref{fig:align_define}) whose position angle
$\phi_\mathrm{sur}$ is given by the relation
\begin{equation}
\tan ( 2 \phi_\mathrm{sur} ) = \frac{2 \mu_{xy}}{\mu_{xx}-\mu_{yy}}
\equiv \beta  .
\end{equation}
With the usual convention that position angle increases from north
through east, choosing the correct branch of the tangent function
yields the position angle
\begin{eqnarray}
\phi_\mathrm{sur} &=& \frac{1}{2} \tan^{-1} \beta
\qquad\qquad \ \ [\mu_{xx} > \mu_{yy} , \mu_{xy} > 0 ] \nonumber \\
\phi_\mathrm{sur} &=&  \frac{1}{2} ( 180\degr + \tan^{-1} \beta )
\quad [ \mu_{xx} < \mu_{yy} ] \\
\phi_\mathrm{sur} &=& \frac{1}{2} ( 360\degr + \tan^{-1} \beta )
\quad [ \mu_{xx} > \mu_{yy} , \mu_{xy} < 0 ]  . \nonumber
\end{eqnarray}
Defined in this way, the position angle lies in the range
$0\degr < \phi_\mathrm{sur} < 180\degr$.

Knowing the position angle $\phi_\mathrm{pa}$ for the target galaxy and
the position angle $\phi_\mathrm{sur}$ for the distribution of surrounding
galaxies, we define the alignment angle $\Phi$ as the angular difference
between $\phi_\mathrm{pa}$ and $\phi_\mathrm{sur}$, constrained to lie in the
interval $0\degr \leq \Phi \leq 90\degr$. Figure~\ref{fig:align_define} shows
how the alignment angle $\Phi$ is defined, using a randomly selected target 
galaxy as an example.

\section{Analysis and Results} 
\label{sec:analysis}

After computing the alignment angle $\Phi$ for the target galaxies in our sample,
we can examine the distribution of $\Phi$ for each of the four subsamples,
divided by  color and luminosity. Table~\ref{align_stats} presents the mean
alignment angle and estimated error in the mean for each subsample of target
galaxies with surrounding galaxies within $r_p = 3.0\, \mathrm{Mpc}$. 
If the alignment angle is randomly distributed, we expect a mean
alignment angle $\langle \Phi \rangle = 45\degr$. In Table~\ref{align_stats},
only the red subsamples show a statistically significant difference from $\langle
\Phi \rangle = 45\degr$. The luminous red (LR) subsample has $\langle \Phi
\rangle < 45\degr$ at the $9.7 \sigma$ level, while the faint red (FR)
subsample also has $\langle \Phi \rangle < 45\degr$, but at the $4.3 \sigma$
level. A mean alignment angle slightly less than $45\degr$ indicates that the
images of the red target galaxies have a slight but statistically significant
tendency to align parallel to the surrounding structure. On the other hand, the
fact that the blue target galaxies have a mean alignment angle indistinguishable
from $45\degr$ does not necessarily imply that they are randomly oriented
relative to the surrounding structure; in a toy model, if half were parallel
($\Phi = 0\degr$) and half were perpendicular ($\Phi = 90\degr$) to the
surrounding structure, that too would yield $\langle \Phi \rangle = 45\degr$.

\begin{deluxetable}{lcc}
\label{align_stats}
\tablecolumns{3}
\tablecaption{Alignment Statistics}
\tablehead{ \colhead{Galaxy sample} & \colhead{$\langle \Phi \rangle$} & \colhead{\# target galaxies} }
\startdata 
    Luminous Blue & $45.140\degr \pm 0.085\degr$ & 92394\\
    Faint Blue & $44.975\degr \pm 0.085\degr$ & 92394\\
    Luminous Red & $44.227\degr \pm 0.080\degr$ & 105965\\
    Faint Red & $44.658\degr \pm 0.080\degr$ & 105965\\
\enddata
\end{deluxetable}

\begin{figure*}[ht!]
\epsscale{1.1}
\plotone{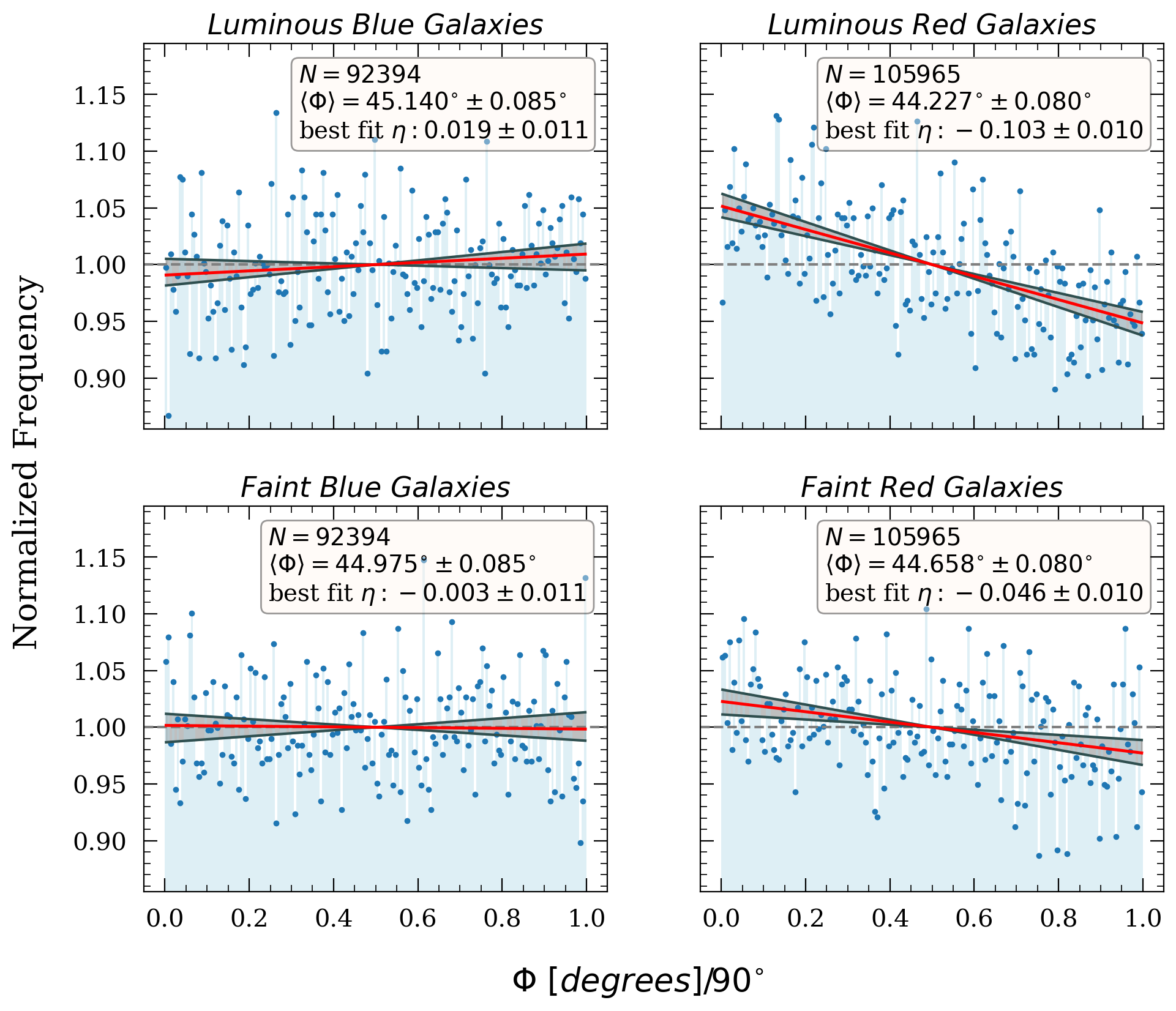}
\caption{Normalized distribution of alignment angle $\Phi$ for the four
subsamples of target galaxies. The best linear fit is shown as the
red line in each panel. The gray lines indicate the range of slopes that
yield $P_\mathrm{ks} > 0.1$ in a Kolmogorov-Smirnov test.}
\label{fig:3}
\end{figure*}

To analyze the results further, we plot a histogram of the distribution of $\Phi$
for each subsample, with bins of width $\Delta\Phi = 0.5\degr$.
Figure~\ref{fig:3} shows the binned distribution of the normalized alignment
angle, $x \equiv \Phi / 90\degr$. To model the distribution function $f(x)$, we
assume a linear fit:
\begin{equation}
f(x) = 1 + \eta ( x - 0.5) .
\end{equation}
In this normalized linear fit, the only variable parameter is the
slope $\eta$.

The best fitting slope $\eta$ and its corresponding $1 \sigma$ error for 
each subsample is found by doing a linear least-squares fit; the resulting 
fits are shown as the red lines in Figure~\ref{fig:3}. In addition, we 
perform non-parametric Kolmogorov-Smirnov (KS) tests, comparing the 
cumulative distribution function for the unbinned data with the cumulative
distribution function for our assumed linear fit,
\begin{equation}
F ( < x ) = x + 0.5 \eta ( x^2 - x ) .
\end{equation}
The gray lines in each panel of Figure~\ref{fig:3} represent the
range in the slope $\eta$ for which the KS test yields a probability
$P_\mathrm{ks} \geq 0.1$. The KS test indicates that both blue subsamples
are consistent with having a random distribution of
alignment angle: the assumption of $\eta = 0$ yields $P_\mathrm{ks} =
0.99$ for the faint blue subsample and $P_\mathrm{ks} = 0.41$ for the
luminous blue subsample. The red subsamples, however, are strongly
inconsistent with having a random distribution of alignment
angle: the assumption of $\eta = 0$ yields $P_\mathrm{ks} = 5
\times 10^{-4}$ for the faint red subsample and $P_\mathrm{ks} = 4
\times 10^{-17}$ for the luminous red subsample.

\begin{figure*}[ht!]
\plotone{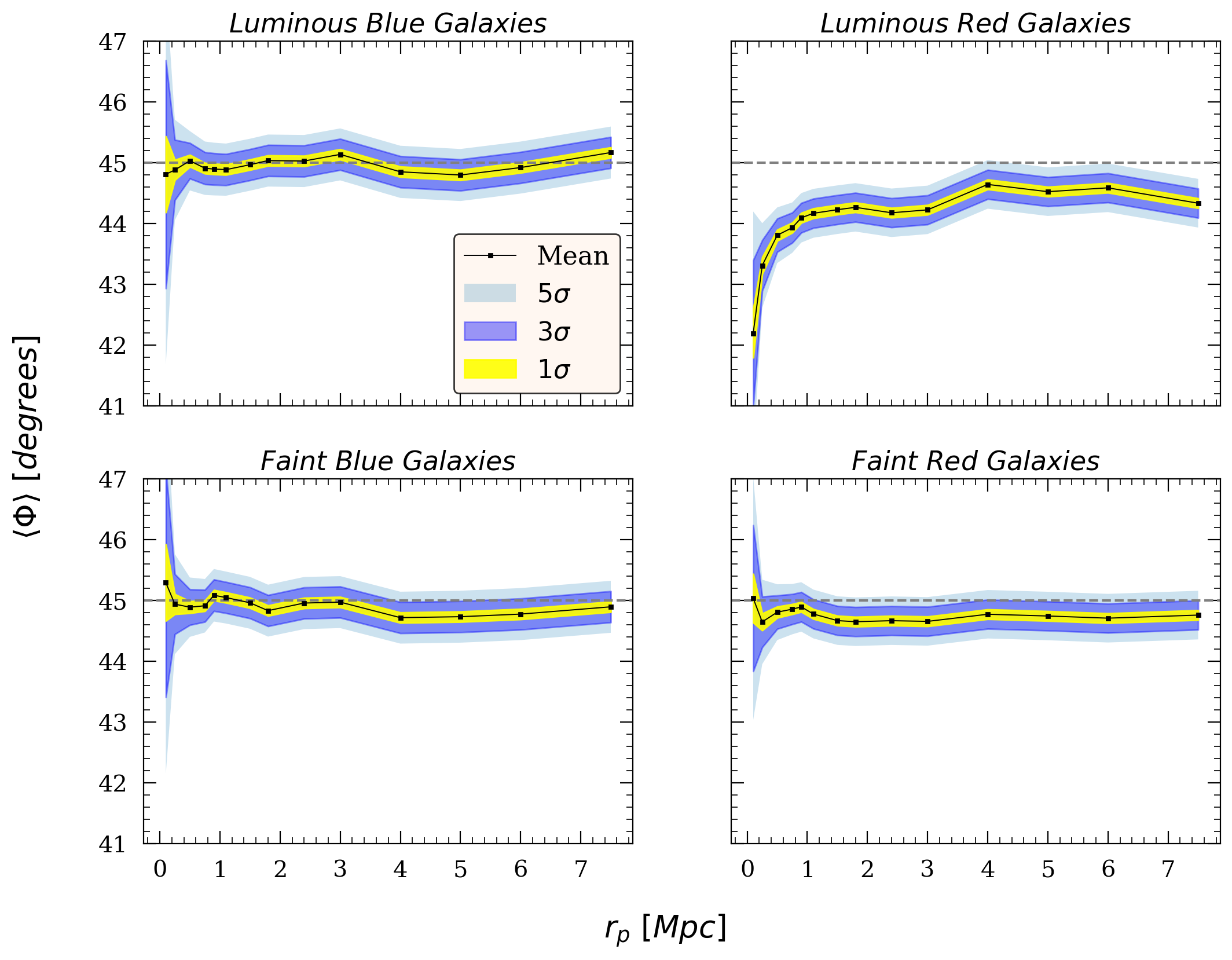}
\caption{Average alignment angle $\langle \Phi \rangle$ as a function of 
maximum projected surrounding radius $r_p = 0.1 \to 7.5\, \mathrm{Mpc}$ for
the four  subsamples of target galaxies.}
\label{fig:4}
\end{figure*}

The above analysis examines the alignment of galaxy images with the
distribution of surrounding galaxies within a projected separation
$r_p = 3.0\, \mathrm{Mpc}$. By varying the limiting projected radius 
$r_p$, we can investigate trends in the alignment angle as a function of
physical scale. Using the SDSS Legacy Survey, we cannot reliably measure
the alignment angle $\Phi$ on scales smaller than $r_p \sim 0.1\,\mathrm{Mpc}$
due to an insufficient number of surrounding galaxies per target galaxy, 
or on scales larger than $r_p \sim 7.5\, \mathrm{Mpc}$ due to the limited 
survey footprint. Given these limitations, in Figure~\ref{fig:4}, we plot 
the average alignment angle when the limiting projected radius lies in 
the range $r_p = 0.1 \to 7.5\,\mathrm{Mpc}$.
In Figure~\ref{fig:4}, the narrowest (yellow) band represents the $1 \sigma$
error interval, the middle (darker blue) band represents the $3 \sigma$ error
interval, and the broadest (pale blue) band represents the $5 \sigma$ error
interval. Throughout the entire range of $r_p$ studied, the blue galaxies
(left panels in Figure~\ref{fig:4}) fail to show a significant difference 
from $\langle \Phi \rangle = 45\degr$ at a significance level $> 3 \sigma$.
By contrast, the luminous red galaxies (upper right panel) have
$\langle \Phi \rangle < 45\degr$ at a significance $> 4.5 \sigma$
throughout the range of $r_p$ with the average alignment becoming more 
parallel for smaller $r_p$. The results for the luminous red galaxies 
are consistent with $\langle \Phi \rangle \approx 44.5\degr$ for the
outermost bins ($r_p \geq 4\,\mathrm{Mpc}$). In the range $1\,\textrm{Mpc} <
r_p < 3\,\textrm{Mpc}$, the mean alignment angle is $\langle \Phi \rangle
\approx 44.2\degr$. At $r_p < 1$\,Mpc, the mean alignment angle decreases
steadily as $r_p$ becomes smaller, reaching $\langle \Phi \rangle =
42.2\degr \pm 0.4\degr$ at $r_p = 0.1$\,Mpc.

The range $1\,\textrm{Mpc} < r_p < 3\,\textrm{Mpc}$ is also where faint
red galaxies (lower right panel of Figure~\ref{fig:4}) show the most statistically
significant alignment (at $>4.3\sigma$). This range of length scales corresponds
to the size of rich clusters of galaxies \citep{2018NewA...58...61B}. On this scale,
we are detecting the alignment of bright galaxies with the projected long axis of
the cluster as a whole \citep{1988AJ.....95..298T,2010MNRAS.405.2023N,2017NatAs...1E.157W}.
On smaller scales, the luminous red galaxies (which include the brightest cluster galaxies)
show a strong parallel alignment with satellite galaxies down to the scale $r_p = 0.1$\,Mpc.
However, the faint red galaxies in our sample show no statistically significant alignment
at $r_p < 1$\,Mpc.

\begin{figure*}[ht!]
\plotone{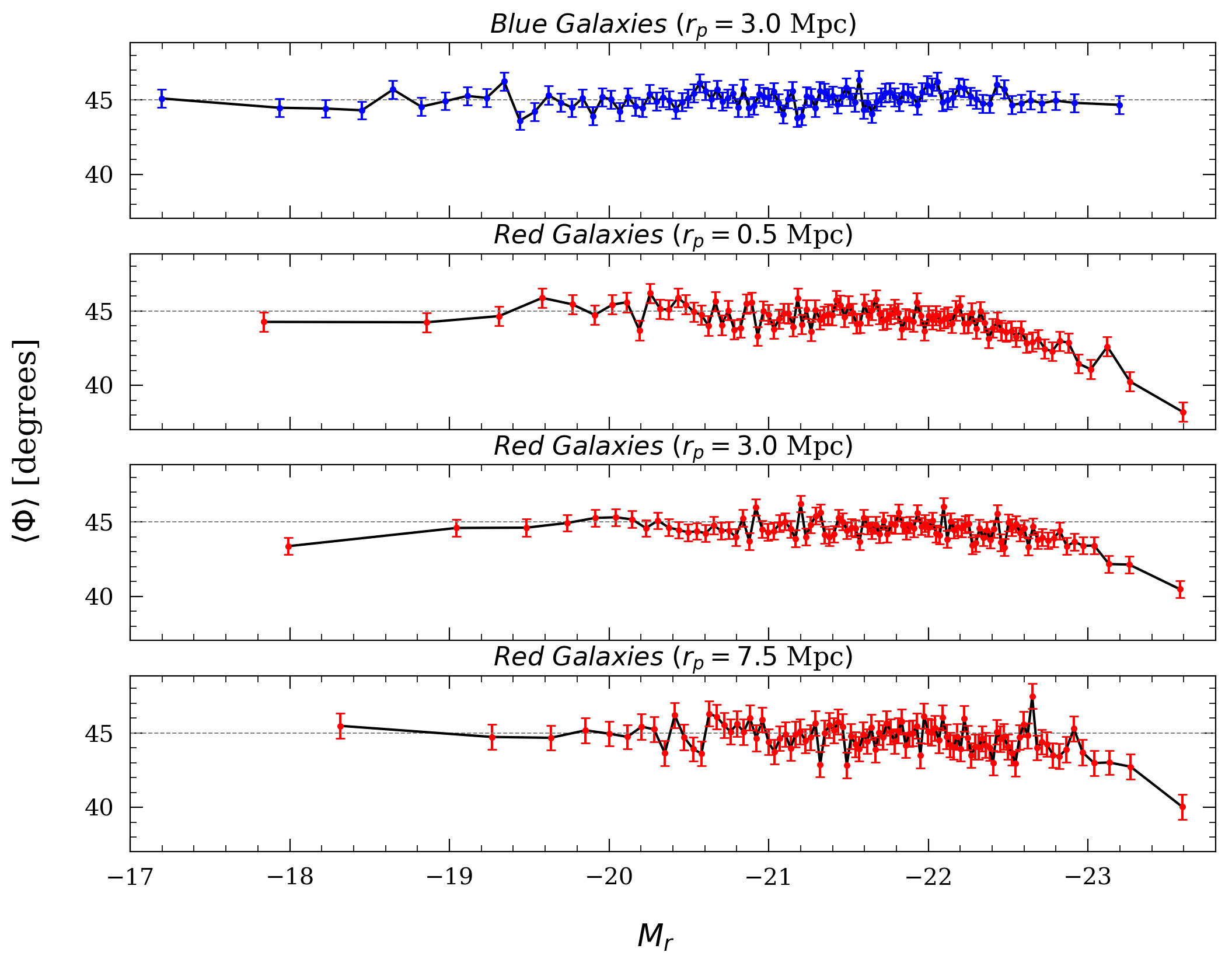}
\caption{Average alignment angle $\langle \Phi \rangle$ as a function of target galaxy absolute magnitudes $M_r$ for blue and red subsamples, using the color divider of \citet{James_2022}. (\textit{a}) blue target galaxies with surrounding galaxies out to $r_p = 3.0\, \mathrm{Mpc}$; (\textit{b}) red target galaxies with surrounding galaxies out to $r_p = 0.5\, \mathrm{Mpc}$; (\textit{c}) red target galaxies with surrounding galaxies out to $r_p = 3.0\, \mathrm{Mpc}$; (\textit{d}) red target galaxies with surrounding galaxies out to $r_p = 7.5\, \mathrm{Mpc}$.}
\label{fig:5}
\end{figure*}

To look in more detail at the dependence of alignment angle on luminosity, we plot
the average alignment angle as a function of the absolute magnitude for the red and
blue samples separately. The data are binned as percentiles, so that each bin
contains the same number of target galaxies and there are 100 bins in total. 
Our large data sample yields 2119 red galaxies per bin, and 1848 blue 
galaxies per bin. The average alignment angles, binned in this manner, 
are plotted in Figure~\ref{fig:5}. The error bars on each point are the 
standard deviation of the bin divided by the square root of the number in 
the bin; that is, the estimated error in the mean.
For blue target galaxies, the average alignment angle is consistent with $45\degr$
(random orientation) throughout the full range of absolute magnitude. Figure~\ref{fig:5}(a)
shows the average alignment angle versus absolute magnitude within $r_p = 3.0$\,Mpc for
blue target galaxies; however, the lack of preferred alignment angle holds over the
entire range from $r_p = 0.1$\,Mpc to $r_p = 7.5$\,Mpc. For the red target galaxies,
there is a clear trend from a preferred parallel alignment at high luminosities to
no preferred alignment at low luminosities. For the most luminous red target galaxies,
the tendency for parallel alignment  is statistically very strong. This holds for all
values of $r_p$ with the average alignment angle following the trend with $r_p$ as shown in
Figure~\ref{fig:4}. Panels (b), (c), and (d) in Figure~\ref{fig:5} show the average
alignment angle versus absolute magnitudes for the red target galaxies for $r_p = 0.5$, 3.0,
and 7.5\,Mpc, respectively. The most luminous 1\% of the red target galaxies ($M_r < -23.35$)
with surrounding radius $r_p = 3.0$\,Mpc, as shown in Figure~\ref{fig:5}(c), have
$\langle \Phi \rangle = 40.49\degr \pm 0.56\degr$; this is less than $45\degr$ at the
$8.1 \sigma$ level. If we bin together  the most luminous 2\% of the red target galaxies
($M_r < -23.19$), we find  that they have $\langle \Phi \rangle = 41.32\degr \pm 0.40\degr$ 
($9.2 \sigma$), and the most luminous 3\% of the red target galaxies ($M_r < -23.08$) have
$\langle \Phi \rangle = 41.60\degr \pm 0.33\degr$ ($10.3 \sigma$).

We performed an additional analysis to determine whether there exists
a correlation between the average alignment angle $\langle \Phi \rangle$
and the number $N$ of surrounding galaxies; this indicates whether galaxies
in higher-density environments tend to have stronger or weaker alignment
signals than those in lower-density environments. We first computed
$\overline{N} (z)$, the mean number of surrounding galaxies within
a projected distance $r_p = 3.0$\,Mpc, as a function of target galaxy redshift.
Splitting each sample into lower-density $\left( N < \overline{N} (z) \right)$ 
and higher-density $\left( N > \overline{N} (z) \right)$ subsamples,
we computed their average alignment angles. We would see a
significant difference between the average alignments of the lower-density
and higher-density subsamples if the alignment depends on the local density. 
The differences between the lower- and higher-density subsamples are consistent with zero
for the luminous blue, faint blue, and faint red galaxy subsamples (see the ``Local density" 
columns in Table~\ref{additional_stats}).
For the luminous red galaxies, the  lower-density 
$\left( \langle \Phi_\mathrm{low} \rangle = 44.33\degr \pm 0.11\degr \right)$ 
and higher-density 
$\left( \langle \Phi_\mathrm{high} \rangle = 44.10\degr \pm 0.12\degr \right)$ 
alignment averages are different only at the $1.4\sigma$ level.

We also tested to see whether the alignment angle $\Phi$ is correlated
with the magnitude of a target galaxy's offset from its surrounding galaxies.
For example, a target galaxy at the center of an elongated filament of
galaxies may have a different alignment signal from a target galaxy at
the end of the filament. The offset of the distribution of surrounding
galaxies relative to the target galaxy is given by the first order moments
$(\mu_x , \mu_y)$ in the Cartesian frame centered on the target galaxy
(equation~\ref{eq:mu_y}). This can be converted to
a fractional offset $f$ by dividing $(\mu_x^2+\mu_y^2)^{1/2}$ by the angular
equivalent of the projected radius limit $r_p = 3.0\, \mathrm{Mpc}$. Using their medians 
$( f_{\mathrm{med,LB}} = 0.10$, $f_{\mathrm{med,FB}} = 0.08$, $f_{\mathrm{med,LR}} = 0.10$, $f_{\mathrm{med,FR}} = 0.09 )$, 
we split each of the four subsamples (LB, FB, LR, and FR) into small offset 
$\left( f < f_{\mathrm{med}} \right)$ and large offset $\left( f > f_{\mathrm{med}} \right)$ 
samples and computed their average alignment angles. No significant 
differences were seen for the luminous blue, faint blue, and faint red 
galaxy samples at $> 0.8\sigma$ level (see the ``Fractional offset" columns in 
Table~\ref{additional_stats}). For the luminous red galaxies, the small offset 
$\left( \langle \Phi_\mathrm{small} \rangle = 44.12\degr \pm 0.11\degr \right)$ 
and large offset 
$\left( \langle \Phi_\mathrm{large} \rangle = 44.34\degr \pm 0.11\degr \right)$ 
alignment averages are different at only the $1.4\sigma$ level.

\begin{deluxetable*}{lcccc}
\label{additional_stats}
\tablecolumns{5}
\tablecaption{Average Alignment Angle $\langle \Phi \rangle$ vs Local Density and Fractional Offset}
\tablehead{ \vspace{0.1cm} & \multicolumn{2}{c}{Local Density} & \multicolumn{2}{c}{Fractional Offset} \\ 
\cline{2-5} 
 & \colhead{Lower density} & \colhead{Higher density} & \colhead{Small offset} & \colhead{Large offset \vspace{-0.3cm}} \\
\colhead{Galaxy sample} & & & & \vspace{-0.4cm} \\
 & \colhead{$\left( N < \overline{N} (z) \right)$} & \colhead{$\left( N > \overline{N} (z) \right)$} & \colhead{$\left( f < f_{med} \right)$} & \colhead{$\left( f > f_{med} \right)$}}
\startdata 
    Luminous Blue & $45.14\degr \pm 0.11\degr$ & $45.14\degr \pm 0.14\degr$ & $45.16\degr \pm 0.12\degr$ & $45.12\degr \pm 0.12\degr$\\
    Faint Blue & $44.98\degr \pm 0.11\degr$ & $44.98\degr \pm 0.14\degr$ & $44.95\degr \pm 0.12\degr$ & $45.00\degr \pm 0.12\degr$\\
    Luminous Red & $44.33\degr \pm 0.11\degr$ & $44.10\degr \pm 0.12\degr$ & $44.12\degr \pm 0.11\degr$ & $44.34\degr \pm 0.11\degr$\\
    Faint Red & $44.67\degr \pm 0.11\degr$ & $44.64\degr \pm 0.12\degr$ & $44.72\degr \pm 0.11\degr$ & $44.60\degr \pm 0.11\degr$\\
\enddata
\end{deluxetable*}

When the fractional offset $f$ is non-zero, we can also define a position angle
$\phi_\mathrm{off}$ of the line segment drawn from the center of the target
galaxy, at $(0,0)$ in the Cartesian frame, to the location of the mean
position offset at $(\mu_x , \mu_y)$. In this way, we can define
a new alignment angle $\Phi_\mathrm{od}$, representing the difference between
the position angle $\phi_\mathrm{pa}$ of the the target galaxy and the position
angle $\phi_\mathrm{off}$ that points toward the center of the surrounding galaxy
distribution. The alignment angle $\Phi_\mathrm{od}$ thus indicates whether a target
galaxy at the fringes of an overdense region tends to point toward the center of
the overdensity. For all four subsamples of target galaxies (LB, FB, LR, and FR), 
the mean value $\langle \Phi_\mathrm{od} \rangle$ is statistically indistinguishable
from $45\degr$, and the distribution of normalized alignment angle
($x \equiv \Phi / 90\degr$) for each subsample is consistent with a random distribution
($\eta = 0$), again verified using the KS test. In addition, we tested for, but did not find,
a dependence of $\langle \Phi_\mathrm{od} \rangle$ on the magnitude of the fractional offset $f$.

\section{Conclusion} 
\label{sec:conclusion}

The Sloan Digital Sky Survey Legacy survey provides a useful database for
looking at the alignment of relatively luminous galaxies with their
surrounding large scale structure, as traced out by other galaxies in the
survey. In our study, we found that highly luminous red galaxies, with $M_r
< -21.78$, have a highly significant tendency for their images to have
their long axes align with the long axis of the surrounding structure; this
tendency is detectable at a $> 4.5 \sigma$ level for a projected length
scale $0.1$\,Mpc $< r_p \leq 7.5$\,Mpc. Fainter red galaxies, with $M_r >
-21.78$, have a tendency to align in the same sense; however, the alignment
signal has significance level $> 4.3 \sigma$ in the range
$1$\,Mpc $ < r_p < 3$\,Mpc. Blue galaxies have no statistically
significant tendency for their images to align with the surrounding
distribution of galaxies.

The observed difference between alignments of red galaxies and blue
galaxies originates in the different physical mechanisms that determine
their alignment in space. Red galaxies are generally elliptical galaxies,
supported by anisotropic velocity dispersion, rather than being rotationally
supported. The linear alignment model for elliptical galaxies
\citep{2001MNRAS.320L...7C} assumes that their shapes are perturbed by
an external tidal field produced by the surrounding density distribution.
Galaxies close to each other in space experience similar tidal fields
and thus have correlated shape distortions \citep{2004PhRvD..70f3526H}.
Since the power spectrum $P(k)$ of cold dark matter density perturbations
extends to small wavenumber $k$, the linear alignment model predicts
alignment of elliptical galaxies on large scales ($\sim$10\,Mpc or more),
consistent with the results of \citet{2009ApJ...694..214O}, and with
our results for red sequence galaxies. By contrast, galaxies at the
luminous end of the blue sequence are generally rotationally supported
disk galaxies. In the tidal torque model for angular momentum acquisition,
galaxy halos are spun up by interaction between their quadrupole
moment and the surrounding tidal field. However, although the tidal
torque model predicts an alignment between the spin axis of a halo
and its surrounding density field, the spin axis of the stellar disk
is often misaligned with that of the halo it inhabits, as a result
of merging subhalos and accretion events
\citep{2014ApJ...785L..15C, 2017MNRAS.472.1163C, 2018ApJ...864...69L}.
Thus, a lack of alignment for disk galaxies with surrounding structure
is not surprising, even on scales $\sim$1\,Mpc.

Our method for quantifying alignment between galaxy images and larger scale
structure was useful over projected length scales from $r_p \sim 0.1$\,Mpc
to $r_p \sim 7.5$\,Mpc using data from the SDSS Legacy survey. Over this
range of scales, running from the size of an individual galaxy and its
satellites to the correlation length for the distribution of galaxy
clusters, a knowledge of intrinsic alignments is important for understanding
the interplay between galaxy evolution and the evolution of large scale structure
in the universe. The same technique that we used to measure alignment angles in
this work can be applied to projections of 3-dimensional numerical simulations,
giving further insight into the evolution of alignment with decreasing redshift,
and the dependence of alignment on the non-gravitational physics involved
in the formation and evolution of galaxies.

\section{Acknowledgements}
\label{sec:acknowledgements}

This project was begun during the Summer Undergraduate Research Program
of the Ohio State University Department of Astronomy, with support from
the Center for Cosmology and AstroParticle Physics (CCAPP). We thank
Derrick James of the Ohio State University School of Earth Sciences for
his assistance with the color dividers.

Funding for the Sloan Digital Sky Survey IV has been provided by the 
Alfred P. Sloan Foundation, the U.S. Department of Energy Office of 
Science, and the Participating Institutions. SDSS-IV acknowledges support and 
resources from the Center for High Performance Computing  at the 
University of Utah. The SDSS website is \url{www.sdss.org}.

SDSS-IV is managed by the Astrophysical Research Consortium 
for the Participating Institutions of the SDSS Collaboration including 
the Brazilian Participation Group, the Carnegie Institution for Science, 
Carnegie Mellon University, Center for Astrophysics | Harvard \& 
Smithsonian, the Chilean Participation Group, the French Participation Group, 
Instituto de Astrof\'isica de Canarias, The Johns Hopkins 
University, Kavli Institute for the Physics and Mathematics of the 
Universe (IPMU) / University of Tokyo, the Korean Participation Group, 
Lawrence Berkeley National Laboratory, Leibniz Institut f\"ur Astrophysik 
Potsdam (AIP),  Max-Planck-Institut f\"ur Astronomie (MPIA Heidelberg), 
Max-Planck-Institut f\"ur Astrophysik (MPA Garching), 
Max-Planck-Institut f\"ur Extraterrestrische Physik (MPE), 
National Astronomical Observatories of China, New Mexico State University, 
New York University, University of Notre Dame, Observat\'ario 
Nacional / MCTI, The Ohio State University, Pennsylvania State 
University, Shanghai Astronomical Observatory, United 
Kingdom Participation Group, Universidad Nacional Aut\'onoma 
de M\'exico, University of Arizona, University of Colorado Boulder, 
University of Oxford, University of Portsmouth, University of Utah, 
University of Virginia, University of Washington, University of 
Wisconsin, Vanderbilt University, and Yale University.

\bibliography{Desai_Ryden}{}
\bibliographystyle{aasjournal}

\end{document}